\pdfoutput=1

\documentclass[twoside,leqno,twocolumn]{article}
\usepackage{ltexpprt}

\usepackage{times}
\usepackage{enumitem}
\usepackage{bm}
\usepackage{xspace}
\usepackage{color}
\usepackage{mathtools}
\usepackage{subcaption}
\usepackage{amsfonts} 
\usepackage{amssymb} 
\usepackage{booktabs}
\usepackage{multirow}
\usepackage{adjustbox}
\usepackage{hyperref}
\usepackage{color,hyperref}
\definecolor{darkblue}{rgb}{0.0,0.0,0.3}
\hypersetup{colorlinks,breaklinks,linkcolor=darkblue,urlcolor=darkblue,anchorcolor=darkblue,citecolor=darkblue}

\setlist[itemize]{leftmargin=12pt}
\setlist[enumerate]{leftmargin=24pt}
\newcommand{\vpara}[1]{\vspace{0.08in}\noindent\textbf{#1 }}
\newcommand{\hide}[1]{} 

\newcommand\hmm[1]{\ifnum\ifhmode\spacefactor\else2000\fi>1000 \uppercase{#1}\else#1\fi}

\newcommand{\ie}{{\sl i.e.\xspace}}

\newcommand{\etal}{{\sl et al.\xspace}}

\newtheorem{thm}{Theorem}[section] 
\newtheorem{definition}[thm]{Definition} 

\newcommand{\mc}{\mathcal}
\newcommand{\mb}{\mathbf}
\newcommand{\mbf}{\mathbf{f}}
\newcommand{\mbg}{\mathbf{g}}
\newcommand{\cX}{\mathcal{X}}

\newcommand{\inc}[1]{\mathrm{Inc}(#1)}

\newcommand{\kenny}{{Carlo}\xspace}
\newcommand{\anglee}{{Steven Spielberg}\xspace}

\newcommand{\elah}{\textsc{AspEm}\xspace}

\begin{document}
\title{\elah: Embedding Learning by Aspects in Heterogeneous Information Networks}

\author{
Yu Shi$^\dagger$\ \ 
Huan Gui$^{\ddagger}$\footnote{The work was done when Huan Gui was a graduate student at UIUC.} \ \ 
Qi Zhu$^\dagger$\ \ 
Lance Kaplan$^\mathsection$\ \ 
Jiawei Han$^\dagger$\\
{\fontsize{10pt}{12pt}\selectfont{\text{$^\dagger$Dept. of Computer Science, University of Illinois at Urbana-Champaign, Urbana, IL USA }}}   \\
{\fontsize{10pt}{12pt}\selectfont{\text{$^\ddagger$Facebook Inc., Menlo Park, CA USA}}}  \ \ \ \ 
{\fontsize{10pt}{12pt}\selectfont{\text{$^\mathsection$U.S. Army Research Laboratory, Adelphi, MD USA}}} \\
{\fontsize{10pt}{12pt}\selectfont{\text{$^\dagger$\{yushi2, qiz3, hanj\}@illinois.edu\ \ \ \ $^\ddagger$huangui@fb.com \ \ \ \ $^\mathsection$lance.m.kaplan.civ@mail.mil}}}\\
}
\date{}

\maketitle


\fancyfoot[R]{\footnotesize{\textbf{Copyright \textcopyright\ 2018 by SIAM\\
Unauthorized reproduction of this article is prohibited}}}







\begin{abstract}
Heterogeneous information networks (HINs) are ubiquitous in real-world applications. 
Due to the heterogeneity in HINs, the typed edges may not fully align with each other. 
In order to capture the semantic subtlety, we propose the concept of aspects with each aspect being a unit representing one underlying semantic facet. 
Meanwhile, network embedding has emerged as a powerful method for learning network representation, where the learned embedding can be used as features in various downstream applications. 
Therefore, we are motivated to propose a novel embedding learning framework---\elah---to preserve the semantic information in HINs based on multiple aspects. 
Instead of preserving information of the network in one semantic space, \elah encapsulates information regarding each aspect individually. 
In order to select aspects for embedding purpose, we further devise a solution for \elah based on dataset-wide statistics.
To corroborate the efficacy of \elah, we conducted experiments on two real-words datasets with two types of applications---classification and link prediction. 
Experiment results demonstrate that \elah can outperform baseline network embedding learning methods by considering multiple aspects, where the aspects can be selected from the given HIN in an unsupervised manner. 
\end{abstract}


\vspace{6pt}


\maketitle


\section{Introduction}\label{sec::introduction}
\begin{figure}[t]
 \centering\includegraphics[width=.9\linewidth]{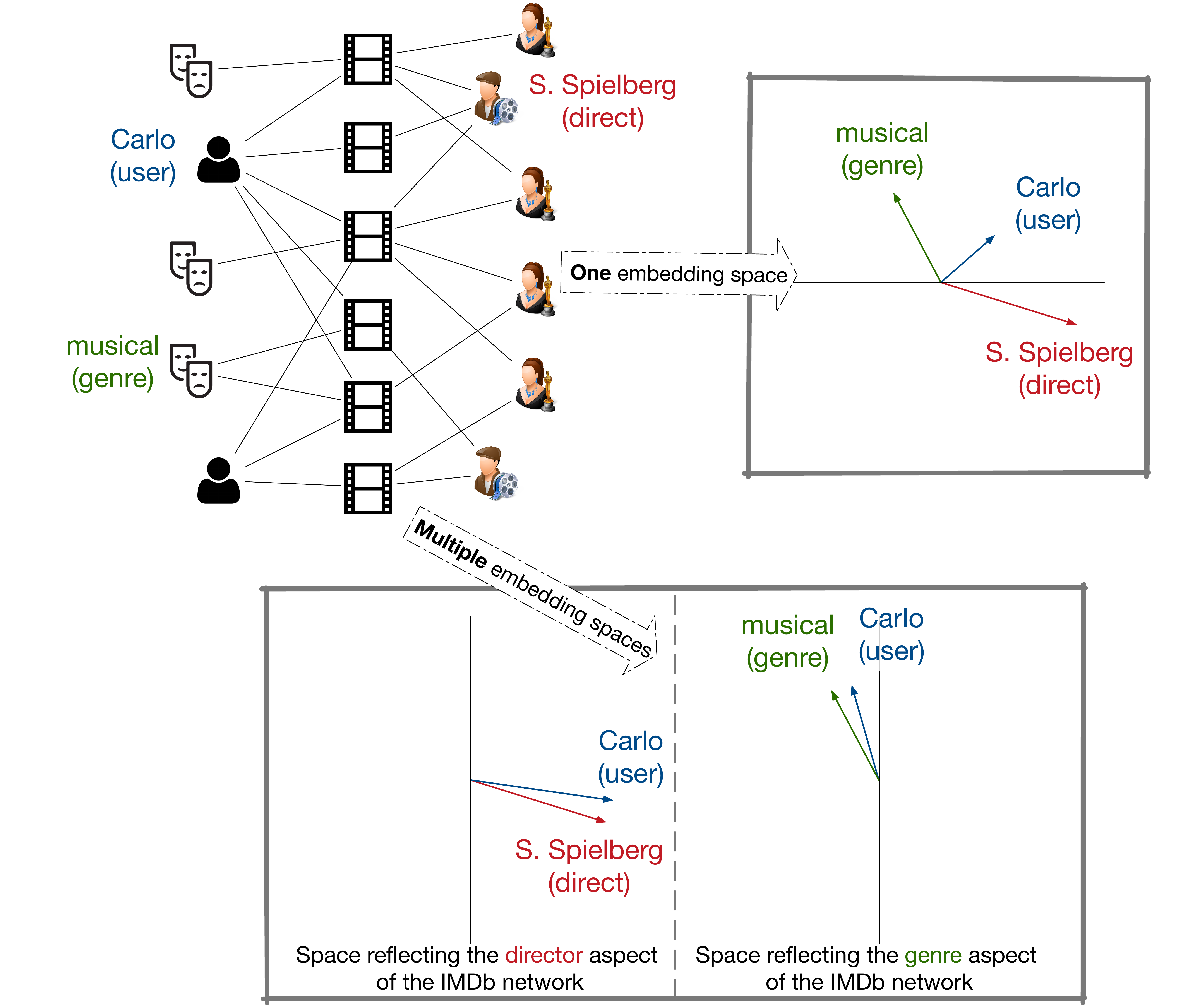}
 \caption[]{
 A toy example of node embeddings in an HIN.  The upper left of the figure depicts the interactions among nodes, where users review movies and movies have various attributes. Carlo likes both musicals and movies directed by S. Spielberg. If all nodes were embedded to one space, Carlo would be close to neither musical nor S. Spielberg due to the dissimilarity between musical and S. Spielberg.  However, by embedding the aspect related to director and that related to genre into separate spaces, Carlo could be close to S. Spielberg in one space, and close to musical in another.
 }\label{fig::intuition}
\end{figure}

In real-world applications, objects of different types interact with each other, forming heterogeneous relations.
Such objects and relations, acting as strongly-typed nodes and edges, constitute numerous \textit{heterogeneous information networks (HINs)}~\cite{shi2017survey, sun2013mining}.
HINs have received increasing interests in the past decade due to its capability of retaining the rich type information, as well as the accompanying wide applications such as recommender system~\cite{yu2014personalized}, clustering~\cite{sun2009ranking}, and outlier detection~\cite{zhuang2014mining}.
As an example, the IMDb network is an HIN containing information about users' preferences over movies and have five different node types: user, movie, actor, director, and genre.

Meanwhile, network embedding has recently emerged as a scalable unsupervised representation learning method~\cite{dong2017metapath2vec, grover2016node2vec, perozzi2014deepwalk, ribeiro2017struc2vec, tang2015pte, tang2015line, wang2016structural}.
In particular, network embedding learning projects the network into low-dimensional space, where each node is represented using a corresponding embedding vector and the relativity among nodes is preserved.
With the semantic information transcribed from the networks, the embedding vectors can be directly used as node features in various downstream applications.
We therefore use the two terms---the embedding of a node and the learned feature of a node---interchangeably in this paper.

The heterogeneity in HINs poses a specific challenge for data mining and applied machine learning.
We hence propose to study the problem of learning embedding in HINs with an emphasis on leveraging the rich and intrinsic type information.
There are multiple attempts in studying HIN embedding or tackling specific application tasks using HIN embedding~\cite{chang2015heterogeneous, chen2017task, gui2016large, tang2015pte}.
Though these studies formulate the problem differently with respective optimization objectives, they share a similar underlining philosophy: using a unified objective function to embed all the nodes into \emph{one} low-dimensional space.

Embedding all the nodes into \emph{one} low-dimensional space, however, may lead to information loss.
Take the IMDb network as example, where users review movies based on their preferences.
Since each movie has several facets, users may review movies with emphasis over different facets.
For instance, both Alice and Bob may like the movie {Star Wars}, but Alice likes it because of Carrie Fisher \textit{(actor)}; while Bob likes it because it is a fantasy movie \textit{(genre)}.
Furthermore, suppose user \kenny likes both movies directed by \anglee and musicals.
Due to the semantic dissimilarity between \anglee and musical, if this HIN were projected into one embedding space as visualized in the upper part of Figure~\ref{fig::intuition}, musical \textit{(genre)} and \anglee \textit{(director)} would be distant from each other, while the user \kenny would be in the middle and close to neither of them.
Therefore, it is of interest to obtain an embedding that can reflect {Carlo}'s preference for both musicals and {Spielberg}'s movies.
To this end, we are motivated to embed the network into two distinct spaces: one for the aspect of genre information whereas the other for that of director information.
In this case, \kenny could be close to musical \textit{(genre)} in the first space and close to \anglee \textit{(director)} in the second space as in the lower part of Figure~\ref{fig::intuition}.

In this paper, we propose a flexible embedding learning framework---\elah---for HINs that mitigates the incompatibility among aspects via considering each aspect separately.
The use of aspects is motivated by the intuition that very distinct relationship could exist between components of a typed network, which has been observed in a special type of HIN \cite{shi2016dynamics}.
Moreover, we demonstrate the feasibility of selecting a set of representative aspects for any HIN using statistics of the network without additional supervision.

It is worth noting that most existing embedding learning methodologies can be extended based on \elah using the principle that different aspects should reside in different embedding spaces.
Therefore, \elah is a principled and flexible framework that has the potential of inheriting the merits of other embedding learning methods.
To the best of our knowledge, this is the first work to study the property of multiple aspects in HIN embedding learning.
Lastly, we summarize our contributions as follows:
\begin{enumerate}
\vspace{-6pt}
\item 
We provide a key insight regarding incompatibility in HINs that each HIN can have multiple aspects that do not align with each other. We thereby identify that embedding algorithms employing only one embedding space may lose subtlety of the given HIN. 
\vspace{-6pt}
\item 
We propose a flexible HIN embedding framework, named \elah, that can mitigate the incompatibility among multiple aspects via considering the semantic information regarding each aspect separately. 
\vspace{-6pt}
\item 
We propose an aspect selection method for \elah, which demonstrates that a set of representative aspects can be selected from any HIN using statistics of the network without additional supervision.
\vspace{-6pt}
\item 
We conduct quantitative experiments on two real-world datasets with various evaluation tasks, which validate the effectiveness of the proposed framework.
\end{enumerate}


\section{Related Work}\label{sec::rel-work}
\vspace{-0.08in}
\vpara{Heterogeneous information networks.}
Heterogeneous information network (HIN) has been extensively studied as a powerful and effective paradigm to model networked data with rich and informative type information~\cite{shi2017survey, sun2013mining}. 
Following this paradigm, a great many applications such as classification, clustering, recommendation, and outlier detection have been studied~\cite{shi2017survey, shi2017prep, sun2013mining, sun2009ranking, yu2014personalized, zhuang2014mining}. 
However, many of these existing works rely on feature engineering~\cite{sun2009ranking, yu2014personalized, zhuang2014mining}.
Meanwhile, we aim at proposing an unsupervised feature learning method for general HINs that can serve as the basis for different downstream applications.

\vpara{Network embedding.}
Network embedding has recently emerged as a representation learning approach for networks~\cite{grover2016node2vec, ou2016asymmetric, perozzi2014deepwalk, ribeiro2017struc2vec, tang2015line, wang2016structural}. 
Unlike traditional unsupervised feature learning approaches~\cite{belkin2001laplacian, roweis2000nonlinear, tenenbaum2000global} that typically arise from the spectral properties of networks, recent advances in network embedding are mostly based on local properties of networks and are therefore more scalable. 
The designs of many homogeneous network embedding algorithms \cite{grover2016node2vec, perozzi2014deepwalk, ribeiro2017struc2vec, tang2015line} trace to the skip-gram model~\cite{mikolov2013distributed} that aims to learn word representations in natural language processing. 
Beyond skip-gram, embedding methods for preserving certain other network properties have also been studied~\cite{ou2016asymmetric, wang2016structural}.

\vpara{Heterogeneous information network embedding.}
There is a line of research on embedding learning for HINs, while the necessity of modeling aspects of an HIN and embedding them into different spaces has been rarely discussed. 
On top of the LINE algorithm~\cite{tang2015line}, Tang et al. propose to learn embedding by traversing all edge types and sampling one edge at a time for each edge type~\cite{tang2015pte}, where the use of type information is shown to be instrumental.
Chang et al. propose to embed HIN with additional node features via deep architectures~\cite{chang2015heterogeneous}, which does not suit for typical HINs consisting of only typed nodes and edges.
Gui et al. devise an HIN embedding algorithm to model a special type of HINs with hyper-edges, which does not apply to general HINs~\cite{gui2016large}.
More recently, an HIN embedding algorithm is proposed, which transcribes semantics in HINs by meta-paths~\cite{dong2017metapath2vec}.
However, this work does not employ multiple embedding spaces for different aspects.
Moreover, it requires the involved meta-paths to be specified as input, while our method is completely unsupervised and can automatically select aspect using statistics of the given HIN.
Embedding in the context of HIN has also been studied to address various application tasks with additional supervision~\cite{chen2017task,  liu2017semantic, pan2016tri, zhang2017triovecevent, zhang2017regions}. 
These methods either yield features specific to given tasks or do not generate node features, and therefore fall outside of the scope of unsupervised HIN embedding that we study.

Additionally, we review the related work on multi-sense embedding in the supplementary file for this paper, which is related but cannot be directly applied to the task of HIN embedding learning with aspects.


\section{Problem Definition}\label{sec::prob-def}
In this section, we formally define the problem of learning embedding from aspects of HINs and related notations. 

\begin{definition}[HIN]
An \textbf{information network} is a directed graph $G = (\mc{V}, \mc{E})$ with a node type mapping $\phi: \mc{V} \rightarrow \mc{T}$ and an edge type mapping $\psi: \mc{E} \rightarrow \mc{R}$. Particularly, when the number of node types $|\mc{T}| > 1$ or the number of edge types $|\mc{R}| > 1$, the network is called a \textbf{heterogeneous information network} (HIN) ~\cite{sun2013mining}.
\end{definition}

In addition, when the network is weighted and directed, we use $W^{(r)}_{uv}$ to denote the weight of an edge $e \in \mc{E}$ with type $r \in \mc{R}$ that goes out from node $u$ and into node $v$.
$D^{O(r)}_u$  and $D^{I(r)}_u$ represent the outward degree of node $u$ (\ie, the sum of weights associated with all edges in type $r$ going outward from $u$) and the inward degree of node $u$ (\ie,  the sum of weights associated with all edges in type $r$ going inward to $u$), respectively. 
For unweighted networks, the degrees can be similarly defined. 
In case a network is undirected, it can be converted to the directed case by simply decomposing every edge to two directed edges with equal weights and opposite directions.

\begin{figure}[t]
  \centering
  \begin{subfigure}[m]{0.52\linewidth}
    \centering\includegraphics[width=\linewidth]{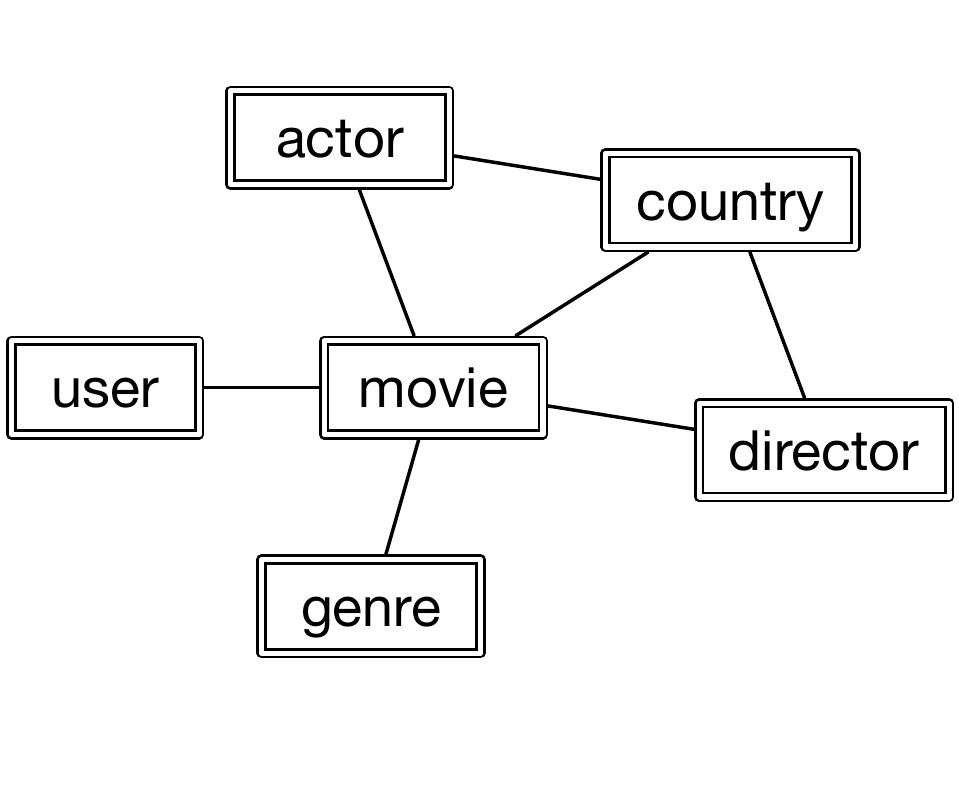}
    \caption{Schema}\label{fig::toy-schema}
    \vspace{-6pt}
  \end{subfigure}
  $\;$
  \begin{subfigure}[m]{0.42\linewidth}
    \centering\includegraphics[width=\linewidth]{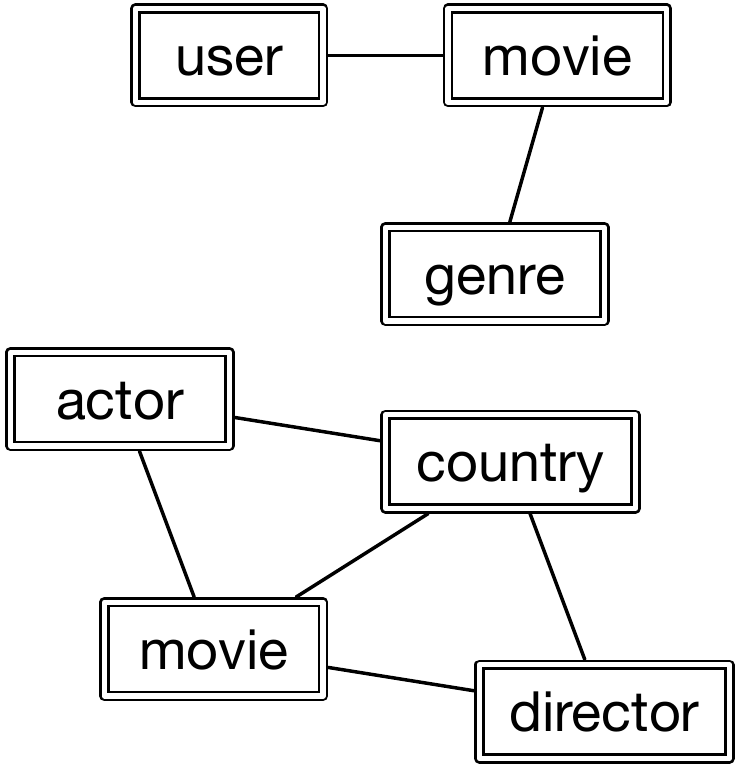}
    \caption{Two aspects}\label{fig::toy-aspect}
    \vspace{-6pt}
  \end{subfigure}
  \caption{The schema and two aspects of an toy HIN with six node types: movie, director, actor, genre, country, and user.}
\end{figure}

Given the typed essence, an HIN can be abstracted using a network schema $\tilde{G} = (\mc{T}, \mc{R})$~\cite{sun2013mining}, which provides meta-information regarding the node types and edge types in the HINs. 
Figure~\ref{fig::toy-schema} gives an example of the schema of a  movie reviewing network as an HIN.

\begin{definition}[Aspect of HIN]\label{def::prob-def}
For a given HIN $G$ with network schema $\tilde{G} = (\mc{T}, \mc{R})$, an \textbf{aspect} of $G$ is defined as a subgraph of the network schema $\tilde{G}$. 
For an aspect $a$, we use $\mc{T}^a \subseteq \mc{T}$ to denote the node types involved in this aspect, and $\mc{R}^a \subseteq \mc{R}$ as the edge types involved in this aspect . 
\end{definition}

As an example, we illustrate two aspects from the schema in Figure~\ref{fig::toy-schema}: one on users' preferences for movies based on genre information (upper part in Figure~\ref{fig::toy-aspect}); and the other on the semantics of movies based on the composite information of directors, actors and their countries (lower part in Figure~\ref{fig::toy-aspect}).
If we denote $\mathcal{A}$ a set of representative aspects generated by a certain method, where information is compatible within each aspect and is not redundant across different aspects, then an HIN with only one aspect will have $|\mathcal{A}| = 1$, $\mc{T}^a = \mc{T}$, and $\mc{R}^a = \mc{R}$.

\hide{
The idea of aspects in HINs is also related to the concept of meta-paths~\cite{sun2011pathsim}. 
Meta paths are defined as structural paths at the meta level, where each meta path consists of a sequence of edge types defined between node types. 
In sharp comparison, aspects are defined as structural relations among different node types at the meta level. 
Naturally, many meta paths can be also modeled as aspects. Moreover, aspects enjoy more flexibility since the structure of each aspect can be arbitrary, besides sequences. Aspects can be used as a tool to represent richer semantic relations among nodes.
Therefore,  aspects in Definition~\ref{def::prob-def} are more versatile.

Now, we are ready to define the problem of embedding learning in HINs from aspects as follows.
}

\begin{definition}[HIN Embedding from Aspects]
Suppose that an HIN $G = (\mc{V}, \mc{E})$ and a set of representative aspects $\mc{A}$ are given. 
For one aspect $a \in \mc{A}$, embedding learning in HIN from one aspect $a$ is to learn a node embedding mapping $f^{a}: \{u \in \mc{V} : \, \phi(u) \in \mc{T}^a\} \rightarrow \mathbb{R}^{d{(a)}}$, where $d{(a)}$ is the embedding dimension for $a$ and $d{(a)} \ll |\mc{V}|$.
For all aspects in $\mc{A}$ and all nodes $u \in \mc{V}$, the problem of \textbf{embedding learning   from aspects in HIN} is to learn corresponding feature vector $\mbf_u$, such that $\mbf_u = \bigoplus_{a \in \mc{A} : \, \phi(u) \in \mc{T}^a} \mbf^{a}_u$, where $\mbf^{a}_u$ is the embedding of node $u$ in aspect $a$.
\end{definition}
We remark that, for nodes of different types, the corresponding $\mbf_u$ might be of different dimensions by definition. 


\section{The \elah Framework}\label{sec::framework}
To address the problem of embedding learning from aspects in HIN, we propose a flexible framework to distinguish the semantic information regarding each aspect. 
Specifically,  for a node $u$, the corresponding embedding vectors  $\mbf^{a}_u$ are inferred independently for different aspects in $\{a \in \mc{A} : \, \phi(u) \in \mc{T}^a\}$.
We name the new framework as \elah, which is short for Aspect Embedding. 
\elah includes three components: (i) selecting a set of representative aspects for the HIN of interest, (ii) learning embedding vectors for each aspect, and (iii) integrating embeddings from multiple aspects. 
We introduce these components as follows.

\subsection{Aspect Selection in HINs}\label{sec::aspect-selection}
Since different aspects are expected to reflect distinct semantic facets of an HIN, an aspect of representative capability should consist of compatible edge types in terms of the information carried by the edges.
Therefore, even without supervision from downstream applications, the incompatibility within each aspect can be leveraged to determine the quality of the aspect, and such incompatibility can be inferred from dataset-wide statistics.

Before introducing the proposed incompatibility measure, $\inc{\cdot}$, we first describe the properties that we posit a proper measure should have as follows.
\begin{property}[Non-negativity]\label{property::non-negativity}
For any aspect $a$, $\inc{a} \geq 0$.
\end{property}
\begin{property}[Monotonicity]\label{property::monotonicity}
For two aspects $a_1$ and $a_2$, if $a_1 \subseteq a_2$, then $\inc{a_1} \leq \inc{a_2}$.
\end{property}
\begin{property}[Convexity]\label{property::convexity}
For two aspects $a_1$ and $a_2$, if their graph intersection has empty edge set, \ie, $\mc{E}(a_1 \cap a_2) = \varnothing$, then $\inc{a_1} + \inc{a_2} \leq \inc{a_1 \cup a_2}$.
\end{property}
We note that the intuition of Property~\ref{property::convexity} is that the incompatibility arises from the co-existence of multiple types of edges.
As a result, generating an aspect by the union of $a_1$ and $a_2$ could only introduce more incompatibility.

To propose our incompatibility measure, we start from the simplest incompatibility-prone scenario: since the incompatibility arises from the co-existence of edge types, the simplest incompatible-prone aspects are those with two edge types joined by a common node type.
In particular, an aspect in this form can be uniquely determined by a schema-level representation $\phi_l \xrightarrow{\psi_l} \phi_c \xrightarrow{\psi_r} \phi_r$,
where $\phi_l, \phi_c, \phi_r \in \mc{T}$ are (not necessarily distinct) node types and $\psi_l, \psi_r \in \mc{R}$ are edge types.
Once the incompatibility measure $\inc{\cdot}$ is defined for this scenario, it can then be generalized to any aspect $a$ by
\begin{equation}\label{eq::inc-aspect}
\inc{a} \coloneqq \sum_{\langle \phi_l, \psi_l, \phi_c, \psi_r, \phi_r \rangle \subseteq a} \inc{\phi_l \xrightarrow{\psi_l} \phi_c \xrightarrow{\psi_r} \phi_r},
\end{equation}
where $\langle \phi_l, \psi_l, \phi_c, \psi_r, \phi_r \rangle \subseteq a$ represents enumerating all such sub-aspects in aspect $a$.
For undirected networks, we do not distinguish $\langle \phi_l, \psi_l, \phi_c, \psi_r, \phi_r \rangle$ and $\langle \phi_r, \psi_r, \phi_c, \psi_l, \phi_l \rangle$ in this enumeration process.
Note that such generalization meets the criteria in Property~\ref{property::monotonicity} and \ref{property::convexity}.

Incompatible edge types result in inconsistent information.
To reflect such intuition, we define the incompatibility measure on aspects of the form $\phi_l \xrightarrow{\psi_l} \phi_c \xrightarrow{\psi_r} \phi_r$ with a Jaccard coefficient--based formulation over each node of type $\phi_c$---the node type that joins two edge types.
Specifically, for node $u$ of type $\phi_c$, we calculate the inconsistency in information observed from $\psi_l$ and $\psi_r$ by
\begin{equation}\label{eq::inc-one-node}
\gamma(u) \coloneqq \frac{\sum_{\phi(\tilde{u}) = \phi_c} \max \big\{ \mb{P}_{u,:}^{\psi_r} (\mb{P}_{\tilde{u},:}^{\psi_r})^\top, \mb{P}_{u,:}^{\psi_l^{-1}} (\mb{P}_{\tilde{u},:}^{\psi_l^{-1}})^\top \big\}}{\sum_{\phi(\tilde{u}) = \phi_c} \min \big\{ \mb{P}_{u,:}^{\psi_r} (\mb{P}_{\tilde{u},:}^{\psi_r})^\top, \mb{P}_{u,:}^{\psi_l^{-1}} (\mb{P}_{\tilde{u},:}^{\psi_l^{-1}})^\top \big\}} - 1,
\end{equation}
where $\mb{M}^{\psi_i}$ is the adjacency matrix of edge type $\psi_i$ and $\mb{P}^{\psi_i}$ is $\mb{M}^{\psi_i}$ after row-wise normalization.
We remark that this formulation, with a difference of minus $1$, is essentially the inverse of Jaccard coefficient over the one-hop neighbors that $u$ can reach via edge type $\psi_l$ and edge type $\psi_r$.
The inverse is taken since greater Jaccard coefficient implies more similarity while we expect more inconsistency, and the minus $1$ is appended so that $\gamma(u) = 0$ when $\mb{P}^{\psi_r} = \mb{P}^{\psi_l^{-1}}$, \ie, no inconsistency if two edge types are identical.
Lastly, we average over all such nodes to find incompatibility score of a simplest incompatible-prone aspect
$$
\inc{\phi_l \xrightarrow{\psi_l} \phi_c \xrightarrow{\psi_r} \phi_r} \coloneqq \frac{1}{|\phi_c^\ast|} \sum_{u \in \phi_c^\ast} \gamma(u),
$$
where $\phi_c^\ast$ is the set of all $u$ in $\phi_c$ such that the denominator in Eq.~\eqref{eq::inc-one-node} is nonzero and $\gamma(u)$ is thereby well-defined.
Note that this definition satisfies Property~\ref{property::non-negativity}.

To select a set $\mc{A}$ of representative aspects for given HIN under any threshold $\theta \in \mathbb{R}_{\geq 0}$, 
(i) an aspect with incompatible score greater than $\theta$ is not eligible to be selected into $\mc{A}$, because it is not semantically consistent enough;
(ii) in case both aspects $a_1$ and $a_2$ have incompatible score below $\theta$ and $a_1 \subset a_2$, we do not select $a_1$ into $\mc{A}$.
We note that the second requirement is intended to keep $\mc{A}$ concise, so that the information across different aspects is not redundant.
Note that when both computation resource and overfitting in downstream application are not of concern, one may explore the potential of trading in model size for gaining additional performance boost by including both $a_1$ and $a_2$ to $\mc{A}$.

We will demonstrate by experiments in Section~\ref{sec::experiment} that this proposed aspect selection method is effective in the sense that (i) \elah built atop this method can outperform baselines that do not model aspects; and (ii) the set of aspects selected using this statistics-based unsupervised method can outperform other comparable sets of aspects.

\subsection{Embedding Learning from One Aspect}\label{sec:sub-embedding-learning}
To design the embedding algorithm for one aspect, we extend the skip-gram model~\cite{mikolov2013distributed} in an approach inspired by existing network embedding studies~\cite{gui2016large, tang2015pte, tang2015line}.
We note that \elah is a flexible framework that can be directly integrated with other homogeneous network embedding methods \cite{grover2016node2vec, perozzi2014deepwalk, ribeiro2017struc2vec, wang2016structural}, other than the adopted skip-gram--based approach, while still enjoying the benefits of modeling aspects in HINs.

For an aspect $a \in \mc{A}$, the associated node embeddings can be denoted as  $\{\mbf^{a}_u\}_{\phi(u) \in \mc{T}^a}$. Recall that $\mc{T}^a$ corresponds to the set of node types included in the aspect $a$.
We model the probability of observing edge $e$ with edge type $r \in \mc{R}^a$ from node $u$ to node $v$ as
\begin{equation}\label{eq::proba}
p^a(v | u, r) = \frac{\exp\big(\mbf^a_u \cdot \mbf^a_v\big)}{\sum_{v' \in \mc{V} : \, \phi(v') = \phi(v)} \exp\big(\mbf^a_u \cdot \mbf^a_{v'}\big)}.
\end{equation}
This equation can be interpreted as the probability of observing $v$ given $u$ and the edge type $r$. 
On the other hand, the empirical conditional probability observed from aspect $a$ is
\begin{equation}\label{eq::emp-proba}
\hat{p}^a(v | u, r) =  {W^{(r)}_{uv}} \big/ {D^{O(r)}_u}.
\end{equation}

To obtain embeddings that reflect the network topology, we seek to minimize the difference between the probability distribution derived from the learned embedding Eq.~\eqref{eq::proba} and the empirical probability distribution observed in data Eq.~\eqref{eq::emp-proba}. Therefore, the embedding learning is reduced to minimizing the following objective function
\begin{equation}\label{eq::obj-1}
\mc{O}^a = \sum_{r \in \mc{R}^a} \sum_{u \in \mc{V}_{O(r)}} \lambda^{(r)}_u d\big(\hat{p}^a(\cdot | u, r), p^a(\cdot | u, r)\big),
\end{equation}
where $\mc{V}_{O(r)} \subseteq \mc{V}$ is the set of all nodes with outgoing type-$r$ edges, $\lambda^{(r)}_u$ is the relative importance of node $u$ in the context of edges with type $r$, and $d(\cdot , \cdot)$ is the KL-divergence. 
Furthermore, we set $\lambda^{(r)}_u \propto D^{O(r)}_u$ with $\lambda^{(r)}_u$ sum up to $1$ for a given edge type $r$. 
Putting pieces together,  Eq.~\eqref{eq::obj-1} can be rewritten as
\begin{equation}\label{eq::obj}
\mc{O}^a = - \sum_{r \in \mc{R}^a} \frac{1}{\Omega^{(r)}} \sum_{u \in \mc{V}_{O(r)}} W^{(r)}_{uv} \log p^a(v | u, r),
\end{equation}
where $\Omega^{(r)} = \sum_{u, v} W^{(r)}_{uv}$.
Consequently, the problem of learning embedding from an aspect $a \in \mc{A}$ is equivalent to solving the following optimization problem
\begin{equation}\label{eq::opt}
\min_{\{\mbf^{a}_u\}_{u : \phi(u) \in \mc{T}^a}} \mc{O}^a.
\end{equation}

With this formulation, information from each aspect of an HIN is transcribed into a different embedding space.

\subsection{Compositing Node Embedding and Edge Embedding}
By solving the optimization problem Eq.~\eqref{eq::opt}, we are able to obtain a feature vector $\mbf^a_u$ for each node $u \in \mc{V}^a$ from the aspect $a \in \mc{A}$, and the final embedding for node $u$ is thereby given by the concatenation of the learned embedding vectors from all aspects involving $u$, \ie, $\mbf_u \coloneqq \bigoplus_{a \in \mc{A} : \, \phi(u) \in \mc{T}^a} \mbf^{a}_u$.
To characterize edges for applications such as link prediction, we follow the method in existing work~\cite{grover2016node2vec} and define the edge embedding mapping $g$ with domain in $\mc{V} \times \mc{V}$ as $g(u, v) = \mbg_{uv} \coloneqq \bigoplus_{a \in \mc{A} : \, \phi(u), \phi(v) \in \mc{T}^a} \mbf^{a}_u \circ \mbf^{a}_v$, where $\circ$ is Hadamard product between two vectors of commensurate dimensions.
We discuss this choice of edge embedding definition in the supplementary file, since it is not the main focus or contribution of our paper.

\subsection{Model Inference}
It is computationally expensive to directly optimize the objective function Eq.~\eqref{eq::obj} since the partition function in Eq.~\eqref{eq::proba} sums over all the nodes in $\mc{V}$.
Therefore, we approximate it with negative sampling~\cite{mikolov2013distributed} and resort to asynchronous stochastic gradient descent (ASGD)~\cite{recht2011hogwild} for optimization as with the common practice in skip-gram--based embedding methods~\cite{grover2016node2vec, perozzi2014deepwalk, tang2015pte, tang2015line}. 
For each iteration in ASGD, we first sample an edge type $r$ from $\mc{R}^a$; then sample an edge $e = (u, v)$ of type $r$ with the sampling probability proportional to $W_{uv}^{(r)}$; and finally obtain negative samples from the noise distribution $P_n^{(r)}(v) \propto \left[ D^{I(r)}_v \right]^{3/4}$~\cite{mikolov2013distributed}. 
The optimization objective for each iteration is therefore
$\log \sigma(\mbf^a_u \cdot \mbf^a_v) + \sum_{i=1}^K \mathbb{E}_{v'_i \sim P_n^{(r)}} \log \sigma(- \mbf^a_u \cdot \mbf^a_{v'_i})$,
where $\sigma(\cdot)$ is the sigmoid function $\sigma(x) = \exp(x)/\big(1+\exp(x)\big)$. 
This optimization procedure shares the same spirit with some existing network embedding algorithms, and one may refer to the network embedding paper by Tang \etal~\cite{tang2015line} for further details.


\section{Experiments}\label{sec::experiment}
In order to provide evidence for the efficacy of \elah, we experiment with two real-world HINs in this section. 
Specifically, the learned embeddings are fed into two types of downstream applications---multi-class classification and link prediction---to answer the following two questions:
\begin{enumerate}
\item[Q1] Does exploiting aspects in HIN embedding learning help better capture the semantics of typed networks in both link prediction and classification tasks?
\item[Q2] Without supervision, is it feasible to select a set of representative aspects just using dataset-wide statistics.
\end{enumerate}

\begin{figure}[t]
  \centering
  \begin{subfigure}[m]{0.48\linewidth}
    \centering\includegraphics[width=\linewidth]{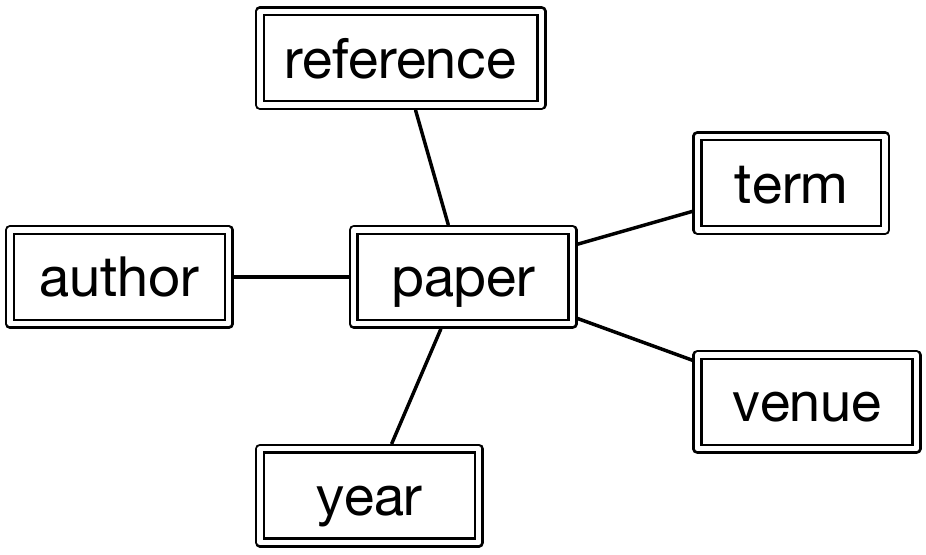}
    \caption{DBLP}\label{fig::dblp-schema}
    \vspace{-6pt}
  \end{subfigure}
  $\;$
  \begin{subfigure}[m]{0.48\linewidth}
    \centering\includegraphics[width=\linewidth]{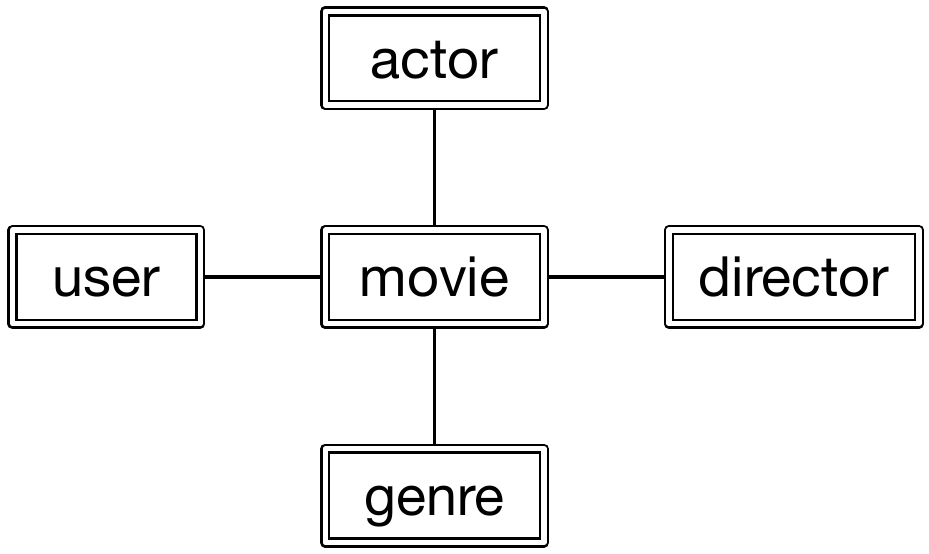}
    \caption{IMDb}\label{fig::imdb-schema}
    \vspace{-6pt}
  \end{subfigure}
  \caption{The network schemas of DBLP and IMDb.}\label{fig::schema}
\end{figure}

\subsection{Data Description}
We use two publicly available real-world HIN datasets: DBLP and IMDb.
\textbf{DBLP} is a bibliographical information network in the computer science domain\footnote{\url{https://aminer.org/citation}}. 
There are six types of nodes in the network: author (A), paper (P), reference (R), term (T), venue (V), and year (Y), where reference corresponds to papers being referred by other papers.
The terms are extracted and released by Chen et al.~\cite{chen2017task}.
The edge types include: author writing paper, paper citing reference, paper containing term, paper publishing in venue, and paper publishing in year.
The corresponding network schema is depicted in Figure~\ref{fig::dblp-schema}.
Note that we distinguish the node type of reference, so that a paper have a different embedding when acting as a reference.
\textbf{IMDb} is an HIN built by linking the movie-attribute information from IMDb and the user-reviewing-movie relationship from MovieLens-100K.\footnote{https://grouplens.org/datasets/movielens/100k/}
There are five types of nodes in the network: user (U), movie (M), actor (A), director (D), and genre (G). 
The edge types include: user reviewing movie, actor featuring in movie, director directing movie, and movie being of genre. 
The network schema can be found in Figure~\ref{fig::imdb-schema}.
We summarize the statistics of the datasets in Table~\ref{tab::data-stat}.
\begin{table}[t]
\centering
    \caption{Basic statistics for the DBLP and IMDb networks.}\label{tab::data-stat}
    \vspace{-9pt}
\scalebox{.75}{
\begin{tabular}{| c | c | c | c | c | c | c |}
    \hline
\multirow{ 2}{* }{DBLP}  & Author & Paper & Reference & Term & Venue & Year \\ \cline{2-7}
  & 1,003,836 & 1,756,680 & 693,406 & 402,687 & 7,528 & 62 \\  \hline
\multirow{ 2}{* }{IMDB}  & User & Movie & Actor & Director & Genre \\ \cline{2-6}
 & 943 & 1,360 & 42,275 & 918 & 23 \\  \cline{1-6}
    \end{tabular}
    }
\end{table}

We use the node types to represent an aspect in these two HINs. 
For example, APY in the DBLP network refers to the aspect involving author, paper, and year, and UMA in IMDb represents the aspect involving user, movie, and actor. 
The schema of each aspect can be easily inferred based on the holistic network schema, as shown in Figure~\ref{fig::schema}.

\subsection{Baseline Methods and Experiment Setting}\label{sec::baseline-and-setting}
To answer Q1 at the beginning of the section, we compare \elah against several unsupervised embedding methods.
\textbf{SVD}~\cite{golub1970singular}: a matrix factorization based method, where singular value decomposition is performed on the adjacent matrix of the homogeneous network and the first $d$ singular vectors are taken as the node embeddings of the network, where $d$ is the dimension of the embedding.
\textbf{DeepWalk}~\cite{perozzi2014deepwalk}: a homogeneous network embedding method, which samples multiple walks starting from each node, and then applies the skip-gram model to learn embedding.
\textbf{LINE}~\cite{tang2015line}: a homogeneous network embedding method, which treats the neighbors of a node as its context, and then applies the skip-gram model to learn embedding. 
\textbf{OneSpace}: as a heterogeneous network embedding method, OneSpace serves as a direct comparison against the proposed \elah algorithm to validate the utility of embedding different aspects into multiple spaces. 
This method is given by the proposed \elah framework with the full HIN schema as the only selected aspect.
We note that the OneSpace method embeds all nodes into only one low-dimensional space. 
In the special case of HINs with star-schema, 
OneSpace is identical to PTE proposed in~\cite{tang2015pte}. 
We remark that DeepWalk is identical to node2vec~\cite{grover2016node2vec} under default hyperparameters. 

For the baselines developed for homogeneous networks, we treat the HIN as a homogeneous network by neglecting the node types.
Additionally, we apply the same downstream learners onto the embeddings yielded by different embedding methods for fair comparison.

\vpara{Parameters.}
While \elah is capable of using different dimensions for different aspects, we employ the same dimension for all aspects out of simplicity.
In other words, we set $d(a_1) = \ldots = d(a_{|\mathcal{A}|}) = d, a_1, \ldots, a_{|\mathcal{A}|} \in \mathcal{A}$. 
In particular, we set $d = 100$ for DBLP and $d = 10$ for IMDb.
For fair comparison, we experiment with two dimensions for every baseline method: (i) the dimension of one aspect used by \elah (\ie, $d$) and (ii) the total dimension of all aspects employed by \elah (\ie, $|\mathcal{A}| \cdot d$). 
We report the better result between the two choices of dimension for every baseline method.
$1,000$ million edges are sampled to learn the embedding on DBLP, and $100$ million edges are sampled on IMDb.
The number of negative samples is set to $5$ following the common practice in network embedding~\cite{tang2015line}.

\vpara{Selected aspects.}
Since all our experiments on DBLP involve the node type author (A), we set the threshold for incompatibility measure $\theta$ to be the \textit{smallest possible value} such that all node types co-exist with the node type author (A) in at least one aspect eligible to be selected to $\mc{A}$ as per the two requirements discussed in Section~\ref{sec::aspect-selection}.
As a result, $\theta$ is set to be $221267$ on DBLP, and the set of selected representative aspects, $\mc{A}$, is \{APRTV, APT\}.
Similarly for IMDb, considering that all its experiments involve the node type user (U), $\theta$ is set to be $1927.68$, and the set of selected representative aspects, $\mc{A}$, is \{UMA, UMD, UMG\}.

The detailed presentation on the calculations and figures involving threshold and aspect selection for both HINs can be found in the supplementary file for this paper.

\subsection{Classification}\label{sec::classification}

For classification tasks, we use the learned embeddings as node features and then classify the nodes into different categories using off-the-shelf classifiers. The classification performance is evaluated using accuracy.
For a set of concerned nodes $\cX$ and node $x \in \cX$, denote $l(x)$ the predicted label of $x$ and denote $l^*(x)$ the ground truth label. 
Then accuracy is defined as 
${\rm Acc.} = \frac{1}{|\cX|} \sum_{x \in cX} \delta\big(l(x) = l^*(x)\big)$, 
where $|\cX|$ is the cardinality of $\cX$ and $\delta(\cdot)$ is the indicator function. 

\begin{table}[t]
\centering
\caption{Classification accuracy in two DBLP tasks.}\label{tab::classification}
\vspace{-9pt}
\resizebox{.4\textwidth}{!}{
\begin{tabular}{c | c | c | c | c}
\toprule \hline
Dataset/task &  \multicolumn{2}{c|}{DBLP-group}  &  \multicolumn{2}{c}{DBLP-area} \\ \hline
Classifier &  LR &  SVM &  LR &  SVM  \\ \hline \hline
SVD & 0.7566 & 0.7550 & 0.8158 & 0.8008  \\ \hline
DeepWalk & 0.6629 & 0.7077 & 0.8308 & 0.8390  \\ \hline
LINE & 0.7037 & 0.7314 & 0.8526 & 0.8540  \\ \hline
OneSpace & 0.7685 & 0.8333 & 0.8758 & 0.8731  \\ \hline \hline
\elah & \textbf{0.8425} & \textbf{0.8889} & \textbf{0.8786} & \textbf{0.8813} \\ \hline
 \bottomrule
\end{tabular}
}
\end{table}

\begin{table*}[t]
\centering
\caption{Link prediction results on DBLP and IMDb.}\label{tab::link-pred}
\vspace{-9pt}
\resizebox{1.\textwidth}{!}{
\begin{tabular}{c | c | c | c | c | c | c | c | c | c | c | c | c }
\toprule \hline
Dataset &  \multicolumn{6}{c|}{DBLP} &  \multicolumn{6}{c}{IMDb}  \\ \hline
Metrics &  $P@1$ & $P@3$ & $P@10$ & $R@1$ & $R@3$ & $R@10$ & $P@1$ & $P@3$ & $P@10$ & $R@1$ & $R@3$ & $R@10$ \\ \hline \hline
SVD & 0.6648 & 0.5164 & 0.2274 & 0.2939 & 0.6178 & 0.8512  &0.2470 & 0.2474 & 0.2249 & 0.0152 & 0.0445 & 0.1343  \\ \hline
DeepWalk & 0.7395 & 0.5297 & 0.2303 & 0.3268 & 0.6329 & 0.8622 & 0.3499 & 0.3605 & 0.3416 & 0.0253 & 0.0774 & 0.2236  \\ \hline
LINE & 0.7404 & 0.5367 & 0.2299 & 0.3267 & 0.6375 & 0.8596 & 0.4782 & 0.4701 & 0.4130 & 0.0379 & 0.1133 & 0.3137 \\ \hline
OneSpace & 0.7440 & 0.5381 & 0.2279 & 0.3301 & 0.6401 & 0.8519 & 0.4665 & 0.4386 & 0.3852 & 0.0435 & 0.1146 & 0.3038 \\ \hline \hline
\elah & \textbf{0.7724} & \textbf{0.5645} & \textbf{0.2356} & \textbf{0.3479} & \textbf{0.6749} & \textbf{0.8810} & \textbf{0.5090} & \textbf{0.4853} & \textbf{0.4219} & \textbf{0.0464} & \textbf{0.1296} & \textbf{0.3420} \\ \hline
 \bottomrule
\end{tabular}
}
\end{table*}

Due to the availability of trustworthy class labels, we perform two classification tasks on DBLP.
The first one (\textbf{DBLP-group}) is on the research group affiliation of each author. 
We consider four research groups led by Christos Faloutsos, Dan Roth, Jiawei Han, and Michael I. Jordan. 
116 authors in the dataset are labeled with such group affiliation. 
The second label set (\textbf{DBLP-area}) is on the primary research area of authors. 
4,040 authors are manually labeled in four research areas: data mining, database, machine learning, and artificial intelligence~\cite{sun2009ranking}.

We experiment with two widely used classifiers. 
One is logistic regression (LR) and  the other is support vector machine (SVM). Both classifiers are based on the liblinear implementation.\footnote{https://www.csie.ntu.edu.tw/~cjlin/liblinear/}
The classification accuracy for different methods are reported in Table~\ref{tab::classification}.

The proposed \elah method outperformed all four baselines in both tasks with either of the two downstream learners applied.
In particular, \elah yielded better results than OneSpace, which confirms our intuition that there exists incompatibility among aspects, and learning node embeddings independently from different aspects can better preserves the semantics of an HIN. 
In addition, we observed that the classification results of \elah were significant better than OneSpace in research group classification; while the improvement of \elah over OneSpace was less significant in research area classification. 
This can be partially explained by that the label of research groups is more relevant to temporal information compared with that of research area, and the presence of the aspect APY in \elah may therefore be more informative for the research group classification task.

Based on the results in Table~\ref{tab::classification}, another observation is that the embedding methods distinguishing node types (OneSpace and \elah) performed better than those not considering node types. 
This observation is in line with previous studies~\cite{gui2016large}, and can be explained by the heterogeneity of node types in HINs.
The nodes of different types in HINs have different properties, such as degrees distribution. 
Simply ignoring such information can lead to information loss.

\subsection{Link Prediction}\label{sec::link-pred}
On experiments with link prediction essence, we perform author identification on the DBLP dataset, and user review prediction on the IMDb dataset.
Precision and recall are used for evaluating these tasks.
Precision at $k$ ($P@k$) is defined as
$P@k = \frac{\# \text{ of true instances at top }k}{k}$,
and recall at $k$ ($R@k$) is defined as
$R@k = \frac{\# \text{ of true instances at top }k}{\# \text{ of total true instances}}$.

We describe the key facts on deriving features for link prediction, and provide further details in the supplementary file.
\textbf{DBLP}---The author identification task on DBLP aims at re-identifying the authors of an anonymized paper, where the reference, term, venue, and year information is still available.
Since papers in the test set do not appear in the training set, their embeddings are hence not available.
Therefore, we use the edge embedding of an author and each attribute of a paper (reference, term, venue, or year) to infer whether this author writes this paper.
Specifically, for both train and test sets, we derive the feature of an author--paper pair by (i) first computing the edge embedding of the concerned author and each attribute of the concerned paper; (ii) then averaging all edge embedding vectors with the same edge type (author--reference, author--term, author--venue, or author--year) to find four edge-type-specific vectors; (iii) finally deriving the feature vector for an author--paper pair by concatenating of the previous four averaged edge embedding vectors.
\textbf{IMDb}---The user review prediction task on IMDb aims at predicting which user reviews a movie.
Features for user--movie pairs are likewise derived as with author--paper pairs in DBLP.

On top of the derived node pair features as well as labels in the training set, logistic regression is trained for inferring the existence of edges in the test set.
We choose the scikit-learn\footnote{http://scikit-learn.org/stable/} implementation with the SAG solver for logistic regression---different from that used for classification---because the SAG solver converges faster than liblinear, and the author identification task on DBLP has a huge number of author--paper pairs as training instances.

From the main results on link prediction presented in Table~\ref{tab::link-pred}, we have observation consistent with the classification tasks that OneSpace and  \elah had better performance than the methods without considering type information. 
Also, \elah outperformed OneSpace. 

\begin{table}[t]
\centering
\caption{Link prediction results ($P@1$) using only one edge. 
}\label{tab::link-pred-one-edge}
\vspace{-9pt}
\resizebox{.48\textwidth}{!}{
\begin{tabular}{c | c | c | c | c   }
\toprule \hline
Edge embedding used & AR & AT & AV & AY \\ \hline \hline
Aspect APRTVY (OneSpace) & 0.6933 & 0.6723 & 0.6501 & 0.3166    \\ \hline \hline
Aspect APRTV   & \textbf{0.7566} & \textbf{0.6977} & \textbf{0.6878} &  ------   \\ \hline
Aspect APR & 0.6071 & ------ & ------ & ------  \\ \hline 
Aspect APT  & ------ & 0.6802 & ------ & ------  \\ \hline 
Aspect APV  & ------ & ------ & 0.5836 & ------  \\ \hline
Aspect APY & ------ & ------ & ------ & 0.3187  \\ \hline
 \bottomrule
\end{tabular}
}
\end{table}

\vpara{Predictive power of single edge embedding.}
In order to better understand the mechanism of \elah in the link prediction tasks, we dissect each aspect and study the predictive power of a single edge embedding from one aspect. 
Specifically, we use each edge embedding over an author-attribute pair from one aspect for link prediction.
Due to space limitation, we focus on the link prediction task on DBLP, because it has the largest number of available labels and can thereby yield most reliable conclusions.
The experimental results are presented in Table~\ref{tab::link-pred-one-edge}, where the rows correspond to the aspect being used for embedding learning and the columns correspond to the edge embedding being used for link prediction.

It can be seen that using the aspect APRTV was better than using the bigger aspect APRTVY for all edge embeddings, where APRTVY was identical to the whole network schema.
Such result provides evidence that for certain HIN datasets, using all the information in the network may be less effective than using partial information (i.e., one aspect). 
We interpret this result as: on the one hand, an author may focus on certain research field that cites certain classic references (R), uses certain terminologies (T), and publishes papers in certain venues (V), \ie, R, T, and V correlate to some extent; on the other hand, an author may be actively publishing papers in a certain range of years (Y). 
However, the information regarding R, T, and V do not align well with Y.
As a result, embedding R, T, V, and Y together into the same space (as in the OneSpace model) led to worse embedding quality even though more types of data were used.
This result further consolidated our insight that HIN can have multiple aspects, and one should embed aspects with different semantics into distinct spaces.

To conclude, the results for classification and link prediction give an affirmative answer to Q1---Distinguishing the information from semantically different aspects can benefit HIN embedding learning. 
 
\subsection{The Impact of Aspect Selection}\label{sec::exp-asp-sel}
In the previous section, we have shown that the aspect selection method proposed in Section~\ref{sec::aspect-selection} can effectively support the \elah framework to outperform embedding methods that do not model aspects in HINs.
In this section, we further address Q2 and demonstrate the set of representative aspects \elah selected using the proposed method is of good quality compared with other selections of aspects.

To this end, we again use the link prediction on DBLP as the downstream evaluation task, and experiment with all sets of aspect that are comparable to \{APRTV, APY\}.
Specifically, each of these comparable sets of aspects (i) has two aspects, and (ii) author and paper appear in both aspects, and other node types exist in exactly one of the two aspects.

\begin{table}[t]
\centering
\caption{Link prediction results using different $2$-combinations aspects on DBLP. 
}\label{tab::asp-sel}
\vspace{-9pt}
\resizebox{.48\textwidth}{!}{
\begin{tabular}{c | c | c | c | c | c | c   }
\toprule \hline
Metrics &  $P@1$ & $P@3$ & $P@10$ & $R@1$ & $R@3$ & $R@10$  \\ \hline \hline
\{APTV, APRY\} & 0.7522 & 0.5476 & 0.2303 & 0.3362 & 0.6524 & 0.8611  \\ \hline
\{APRV, APTY\} & 0.7347 & 0.5327 & 0.2257 & 0.3271 & 0.6327 & 0.8425  \\ \hline
\{APRT, APVY\} & 0.7579 & 0.5556 & 0.2332 & 0.3385 & 0.6614 & 0.8708  \\ \hline
\{APTVY, APR\} & 0.7384 & 0.5360 & 0.2277 & 0.3280 & 0.6372 & 0.8499  \\ \hline
\{APRVY, APT\} & 0.7353 & 0.5356 & 0.2271 & 0.3263 & 0.6355 & 0.8474  \\ \hline
\{APRTY, APV\} & 0.7366 & 0.5362 & 0.2277 & 0.3274 & 0.6364 & 0.8492  \\ \hline \hline
\{APRTV, APY\} & \textbf{0.7724} & \textbf{0.5645} & \textbf{0.2356} & \textbf{0.3479} & \textbf{0.6749} & \textbf{0.8810}    \\ \hline 
 \bottomrule
\end{tabular}
}
\end{table}

From the results presented in Table~\ref{tab::asp-sel}, it can be seen that the set of representative aspects selected by our proposed method, \{APRTV, APY\}, achieved the best performance among all comparable aspect selections.
Note that all the $6$ inferior sets of aspects have inconsistency score, $\inc{\cdot}$, greater than the threshold we set, which can be verified from the numbers provided in the supplementary file.
This result further consolidates the feasibility of selecting representative aspects for any HIN solely by dataset-wide statistics without the need of additional task-specific supervision.

\subsection{Hyperparameter Study}\label{sec::param-sens}
We vary two hyperparameters, one at each time, that play important roles in embedding learning: dimension of embedding spaces and the number of edges sampled in the training phase.
All other parameters are set following Section~\ref{sec::baseline-and-setting}.

\begin{figure}[t]
\begin{minipage}[c][][t]{.24\textwidth}
  \centering
  \includegraphics[width=\textwidth]{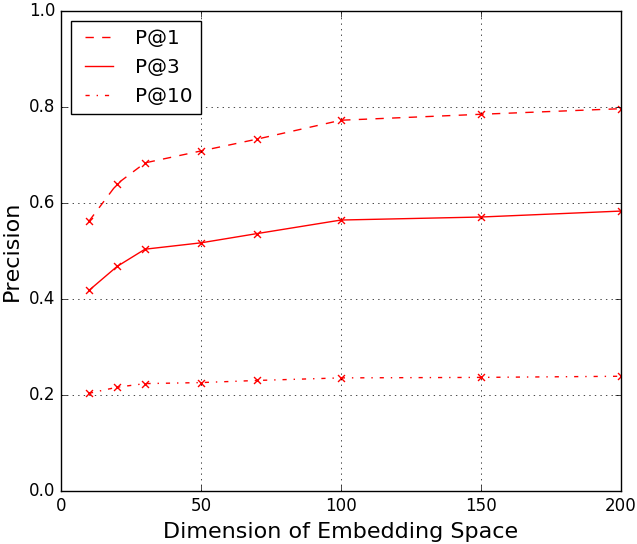}
  \vspace{-12pt}
  \subcaption{}
  \label{fig::dim-precision}
  \par
  \includegraphics[width=\textwidth]{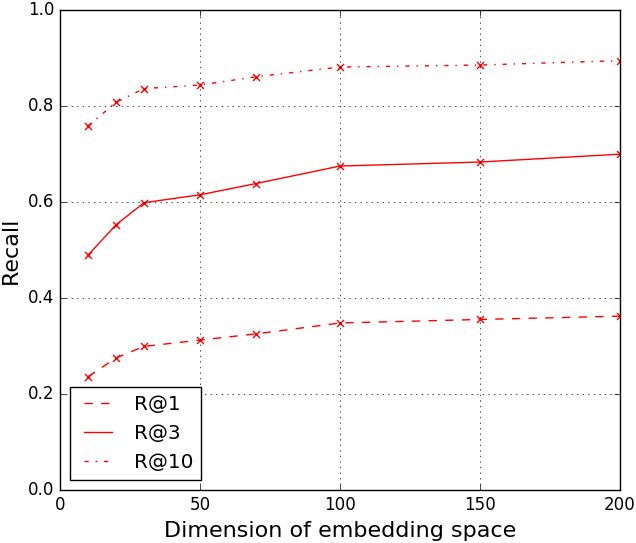}
  \vspace{-12pt}
  \subcaption{}
  \label{fig::dim-recall}
  
\end{minipage}
\begin{minipage}[c][][t]{.24\textwidth}
  \centering
  \includegraphics[width=\textwidth]{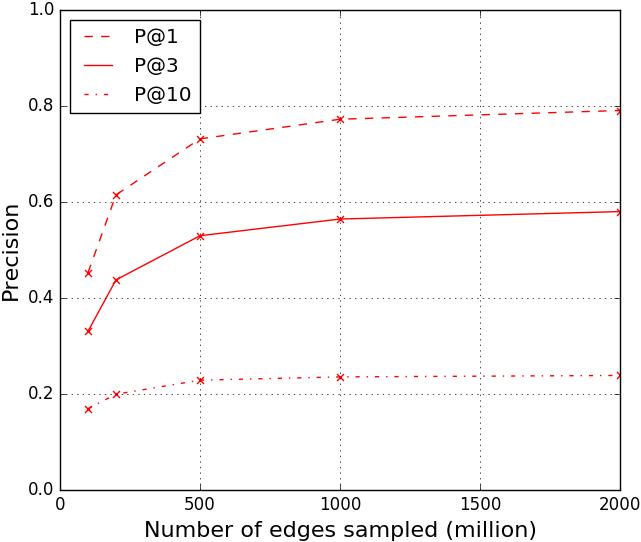}
  \vspace{-12pt}
  \subcaption{}
  \label{fig::nsmp-precision}
  \par
  \includegraphics[width=\textwidth]{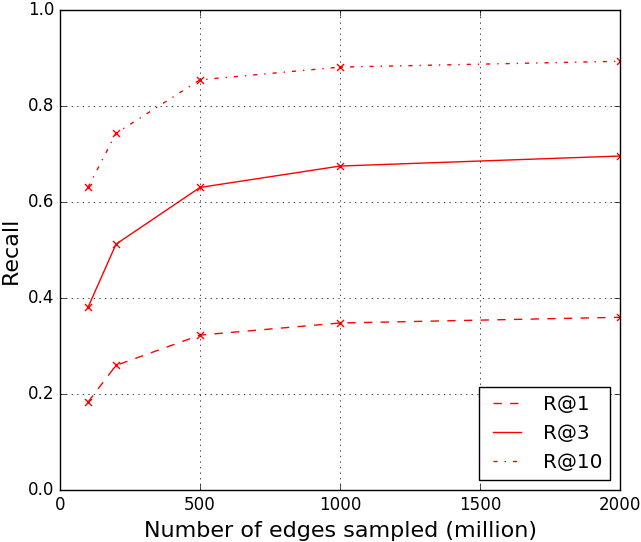}
  \vspace{-12pt}
  \subcaption{}
  \label{fig::nsmp-recall}
\end{minipage}
\caption{(a) and (b) depict the precision and recall against various dimensions employed for the embedding space. (c) and (d) give the precision and recall against various choices of sampled edge numbers.}\label{fig::param-sens}
\end{figure}

The performance in the link prediction task on DBLP is presented in Figure~\ref{fig::param-sens}.
It can be seen that model performance tended to be better as either the dimension of embedding spaces or the number of edges sampled grew, and the growth became less drastic after dimension reached $100$ and number of edges sampled reached $1000$ million. 
Such a pattern agrees with the results in other similar studies~\cite{grover2016node2vec, gui2016large, tang2015line}.


\section{Conclusions and Future Work}\label{sec::conclusion}
In this paper, we study the problem of embedding learning in HINs.
Particularly, we make the key observation that there are multiple aspects in heterogeneous information networks and there might be incompatibility among different aspects.
Therefore, we take advantage of the information encapsulated in each aspect and propose \elah---a new embedding learning framework from aspects, which comes with an unsupervised method to select a set of representative aspects from an HIN.
We conducted experiments to corroborate the efficacy of \elah in better representing the semantic information in HINs. 

To focus on the utility of aspects in HIN embedding, \elah is designed to be simple and flexible with each aspect embedded independently. 
For future work, one may explore optimizing the embeddings for all the aspects jointly, in hope of preserving more intrinsic information among nodes and further boost performance in downstream applications.
Additionally, it is of interest to investigate into aspect selection methods when supervision is further provided.
 
\vpara{Acknowledgments.}
This work was sponsored in part by U.S. Army Research Lab. under Cooperative Agreement No. W911NF-09-2-0053 (NSCTA), DARPA under Agreement No. W911NF-17-C-0099, NSF IIS 16-18481, IIS 17-04532, and IIS-17-41317, and grant 1U54GM114838 awarded by NIGMS through funds provided by the trans-NIH Big Data to Knowledge (BD2K) initiative (www.bd2k.nih.gov). 
The views and conclusions contained in this paper are those of the authors and should not be interpreted as representing any funding agencies.
The U.S. Government is authorized to reproduce and distribute reprints for Government purposes notwithstanding any copyright notation hereon.

\bibliographystyle{siam}
\bibliography{sigproc}

\section*{Supplementary Materials}

\subsection*{Related Work on Multi-Sense Embedding}
The idea of multiple aspects is in a way related to the polysemy of words. 
There have been some studies on inferring multi-sense embeddings of words~\cite{arora2016linear, jauhar2015ontologically, neelakantan2015efficient, vsuster2016bilingual}, which aims at inferring multiple embedding vectors for each word.
However, the two tasks differ significantly in the following perspectives. Firstly, each node may have multiple embeddings due to the semantic subtlety associated with each aspect; while in multi-sense word embedding learning, the number of senses for each word varies. Secondly, we aim at studying the embedding in HINs; while multi-sense embeddings word embedding learning is for textual data.
Therefore, the methods developed for multi-sense embedding learning cannot be directly applied to the task of HIN embedding learning with aspects. 

\subsection*{Discussion on Compositing Edge Embedding}
Instead of simply focusing on the node embeddings, another important component of networks is the interactions among nodes, \ie, edges. 
Characterizing edges is important for downstream applications such as link prediction, which aims to predict whether there is an edge between a pair of nodes for a certain edge type.
Therefore, it is of interest to define the embedding for edges.
In this paper, we simply refer to a function of embeddings of a node pair as edge embedding, even if there might be no edge between the given node pair. 
The function of the edge embeddings is a hyperparameter and can be chosen by various designs. 

Multiple possible ways exist in building edge embedding from the embedding vectors of the two involved nodes. 
In the \elah framework, we bridge node embedding and edge embedding by Hadamard product~\cite{horn1990hadamard}.
We adopt Hadamard product in this design for two reasons:
(i) For a pair of nodes, the inner product of the node embeddings is equivalent to the sum of Hadmard product of the two embeddings. As formulated in Eq.~(4.3), the inner product of the node embeddings plays a vital role in modeling the proximity of edges between nodes. 
(ii) Empirical experiments on three datasets from a previous study~\cite{grover2016node2vec} show that Hadamard product is a choice superior to other options in constructing edge embeddings from node embeddings. 
Specifically, we define the edge embedding mapping $g$ with domain in $\mc{V} \times \mc{V}$ as $g(u, v) = \mbg_{uv} \coloneqq \bigoplus_{a \in \mc{A} : \, \phi(u), \phi(v) \in \mc{T}^a} \mbf^{a}_u \circ \mbf^{a}_v$, where $\circ$ is Hadamard product between two vectors of commensurate dimensions.

We additionally remark that a recent paper~\cite{abu2017learning} specifically addresses the problem of learning edge representation, and defines edge embedding as a parametric function over node embeddings, which is learned from the dataset.
Since the focus of our paper is not to tackle the problem of edge embedding, we simply adopt the aforementioned straightforward Hadamard approach.

\subsection*{Additional Details on Link Prediction Feature Derivation}
We provide further details on deriving features for link prediction tasks on both DBLP and IMDb in addition to the information available in Section~5.1 from the main content of the paper.
\textbf{DBLP}---We randomly selected 32,488 papers into the test set, and take the rest as training data.
Following the procedure proposed by Chen et al.~\cite{chen2017task}, for each paper in test, we randomly sample a set of negative authors, which together with all the true authors constitute the candidate author set of size $100$.
\textbf{IMDb}---As with the DBLP author identification task, we sampled a candidate set of $100$ movies for each user for testing on DBLP.

\subsection*{Incompatibility Score of Each Aspect in DBLP and IMDb}
In this section of the supplementary material, we provide the sufficient statistics for calculating incompatibility of each aspect as defined in Section~4.1 from the main content of the paper.
That is, we provide the incompatibility of aspects of the form $\phi_l \xrightarrow{\psi_l} \phi_c \xrightarrow{\psi_r} \phi_r$ as in Table~\ref{tab::inc-dblp} for DBLP and Table~\ref{tab::inc-imdb} for IMDb.
Note that the proposed \elah framework selects a set of representative aspects $\mc{A}$ for embedding purpose based on their incompatibility, which will be illustrated in the next section.

\begin{table}[!h]
\centering
\caption{Sufficient statistics for incompability on DBLP.}\label{tab::inc-dblp}
\resizebox{.3\textwidth}{!}{
\begin{tabular}{c | c }
\toprule \hline
Aspect & Incompatibility score \\ \hline \hline
$R-P-Y$ & 52753.6 \\ \hline
$A-P-Y$ & 221267. \\ \hline
$T-P-Y$ & 10254.4 \\ \hline
$V-P-Y$ & 1830.08 \\ \hline
$A-P-R$ & 307.988 \\ \hline
$T-P-R$ & 6060.62 \\ \hline
$V-P-R$ & 948.654 \\ \hline
$T-P-A$ & 11518.2 \\ \hline
$V-P-A$ & 5724.80 \\ \hline
$V-P-T$ & 3579.59 \\ \hline
 \bottomrule
\end{tabular}
}
\end{table}

\begin{table}[!h]
\centering
\caption{Sufficient statistics for incompability on IMDb.}\label{tab::inc-imdb}
\resizebox{.3\textwidth}{!}{
\begin{tabular}{c | c }
\toprule \hline
Aspect & Incompatibility score \\ \hline \hline
$U-M-A$ & 171.607 \\ \hline
$D-M-A$ & 1689.76 \\ \hline
$G-M-A$ & 12956.6 \\ \hline
$D-M-U$ & 1927.68 \\ \hline
$G-M-U$ & 636.442 \\ \hline
$G-M-D$ & 531.266 \\ \hline
 \bottomrule
\end{tabular}
}
\end{table}

\subsection*{Aspect Selection in DBLP and IMDb}

\begin{figure*}[th]
 \centering\includegraphics[width=\linewidth]{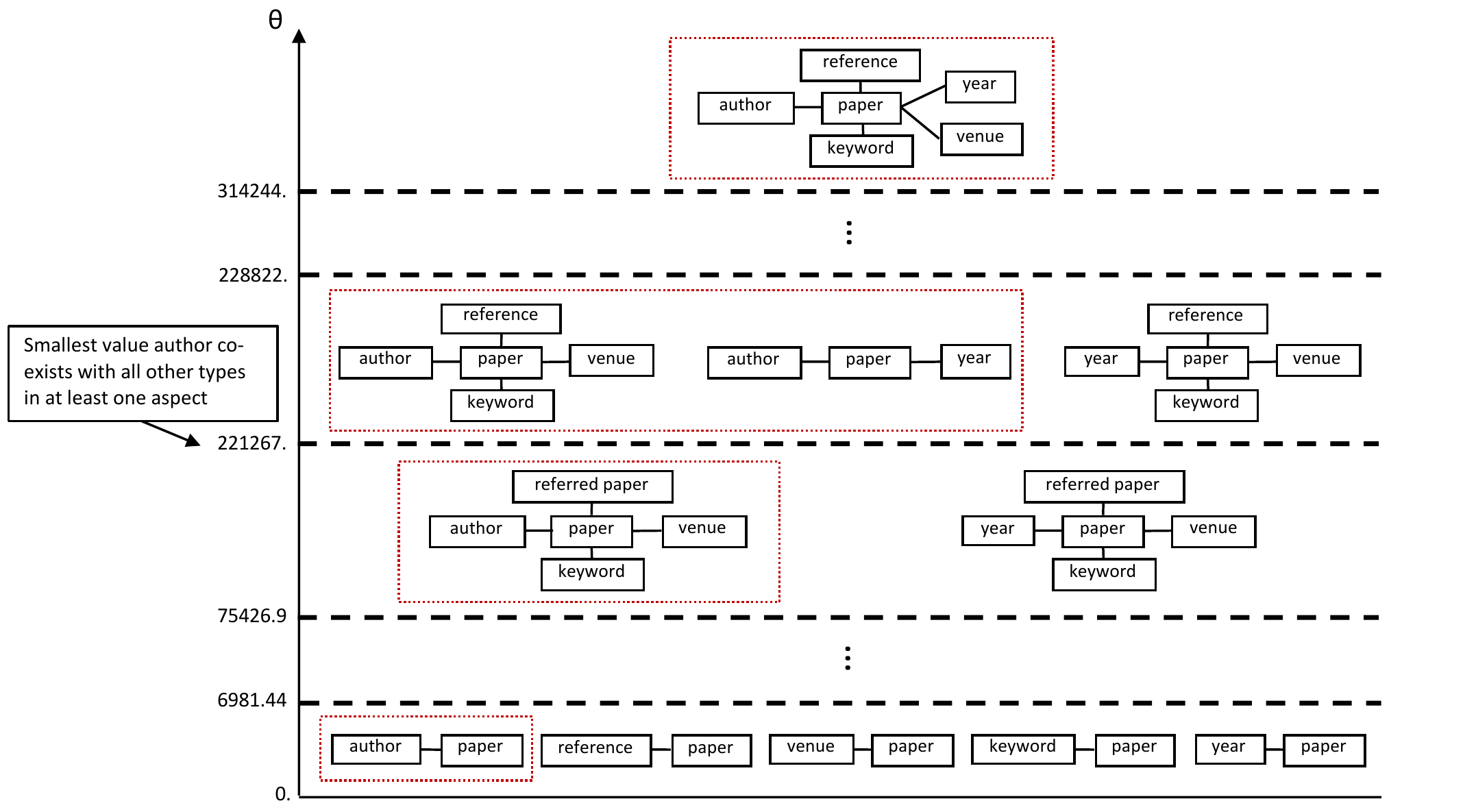}
 \caption{Aspects in DBLP satisfying the two requirements at various threshold $\theta$.}\label{fig::asp-sel-dblp}
\end{figure*}

Using Eq.~(4.1) and the sufficient statistics provide in Table~\ref{tab::inc-dblp} and \ref{tab::inc-imdb}, one can calculate the incompatibility score of any aspect in DBLP and IMDb. 
We proceed to illustrate the aspect selection results using DBLP as example.

Given any threshold $\theta \in \mathbb{R}_{\geq 0}$, (i) any aspect with incompatible score greater than $\theta$ is not eligible to be selected into $\mc{A}$, because it is not meaningful and semantically consistent enough to be one representative aspect of the involved HIN;
(ii) in case both aspects $a_1$ and $a_2$ have incompatible score below $\theta$ and $a_1 \subset a_2$, we do not select $a_1$ into $\mc{A}$.
We note that the second requirement is intended to keep $\mc{A}$ concise and representative in the aspect selection process. 
However, when both computation resource and overfitting in downstream application are not of concern, one may explore the possibility of gaining additional performance boost by adding both $a_1$ and $a_2$ to $\mc{A}$.

Aspects in DBLP satisfying the aforementioned two requirements at various threshold $\theta$ are presented in Figure~\ref{fig::asp-sel-dblp}.
Since all our experiments on DBLP involve the node type author (A), we set $\theta$ to be the \textit{smallest possible value} such that all node types co-exist with the node type author (A) in at least one aspect eligible to be selected to $\mc{A}$ as per the aforementioned two requirements.
Therefore, $\theta$ is set to be $221267$ on DBLP.
One can verify this by calculating $\inc{APRTV} = \inc{APR} + \inc{TPA} + \inc{VPA} + \inc{TPR} + \inc{VPR} + \inc{VPT} = 28139.9$, $\inc{APY} = 221267$, and $\inc{YPRTV} = \inc{RPY} + \inc{TPY} + \inc{VPY} + \inc{TPR} + \inc{VPR} + \inc{VPT} = 75426.9$.

Furthermore, aspects not involving author (A) are additionally excluded from $\mc{A}$ (those outside of the dotted boxes in Figure~\ref{fig::asp-sel-dblp}), because whether or not adding them to $\mc{A}$ does not affect the downstream evaluations.
As a result, the set of selected representative aspects, $\mc{A}$, for DBLP is \{APRTV, APT\}.

Similarly for IMDb, following the same requirements and the consideration that all its experiments involve the node type user (U), $\theta$ is set to be $1927.68$, and the set of selected representative aspects, $\mc{A}$ is \{UMA, UMD, UMG\}.


\end{document}